# A Metrics Based Model for Understandability Quantification

### Mohd Nazir, Raees A. Khan and Khurram Mustafa

**Abstract**— Software developers and maintainers need to read and understand source programs and other software artifacts. The increase in size and complexity of software drastically affects several quality attributes, especially understandability and maintainability. False interpretation often leads to ambiguities, misunderstanding and hence to faulty development results. Despite the fact that software understandability is vital and one of the most significant components of the software development process, it is poorly managed. This is mainly due to the lack of its proper management and control. The paper highlights the importance of understandability in general and as a factor of software testability. Two major contributions are made in the paper. A relation between testability factors and object oriented characteristics has been established as a first contribution. In second contribution, a model has been proposed for estimating understandability of object oriented software using design metrics. In addition, the proposed model has been validated using experimental try-out.

**Index Terms**— Software Quality, Software Design, Software Testability, Testability Factors, Understandability.

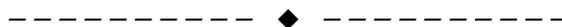

## 1 INTRODUCTION

Companies in the software industry now exist in an environment more turbulent than ever before. To be able to compete in the fast changing environments, companies have to develop smart processes that are easily adapted to the changing environment. Unfortunately, most of the software industries not only fail to deliver a quality product to their customers, but also sometimes they do not understand the relevant quality attributes [11]. Furthermore, in IT industry, schedules are often tightly restricted because of the consumer pressure; developers are forced to weigh the importance of quality against the possibility of missing deadlines. For meeting the target, 'on time delivery', testing time is generally reduced. It increases the potential for defects, leading to problems with the software that include incomplete design, poor quality, high maintenance cost and also the risk of loosing customer satisfaction.

In order to meet the changing requirements of customer or due to many other reasons, software needs to be changed or modified from time to time [10]. This process of modification or maintenance is usually carried out by programmers, which may not have developed that software. They need to read and understand source programs and other relevant documents. Even for the developers of the system, after a gap of few years, it may not be an easy task for them as they themselves might have forgotten the intricacies of the software. False interpretations can lead to misunderstandings and to faulty development results. Without an understanding and the ability to articulate the processes in use, it is not likely that they can be managed and improved. Therefore, the readability and understandability of the software has a lot of influence on the factors that directly or indirectly affect software quality. Complex design may lead to poor testability, which in turn leads to ineffective testing that may result to severe penalties and consequences. It is well understood fact that flaws of design structure have a strong negative impact on quality attributes. But, structuring a high-quality design continues to be an inadequately defined process [14]. Therefore, software design should be built in such a way so as to make them easily understandable, testable, alterable, and preferably stable.

Rest of the paper is organized as follows. In section 2 software testability is addressed. Section 3 briefly describes the identified factors of testability. Section 4 gives an overview of OO characteristics. Section 5 shows the mapping between Testability Factors & OO Characteristics. Section 6 & 7 respectively highlights the significance of understandability and presents a model for its estimation. Empirical validation is given in section 8. Finally, section 9 concludes the paper.

## 2 SOFTWARE TESTABILITY

Testability is one of the most important quality indicators; its measurement leads to the prospects of facilitating and improving a test process. The insight provided by software testability is valuable during design, coding, testing, and quality assurance [3]. The characteristics of testable software like adequate complexity, low coupling and good separation of concerns make it easier for reviewers

————————————————

- *Mohd Nazir is associated with the Department of Computer Science, Jamia Millia Islamia (A Central University), New Delhi, India.*
- *Raees A. Khan is associated with the Department of IT, Babasaheb Bhimrao Ambedkar University, Lucknow, India.*
- *Khurram Mustafa is associated with the Department of Computer Science, Jamia Millia Islamia (A Central University), New Delhi, India.*

to understand the software artifacts under review [9]. Testability results from good Software Engineering practice and an effective software process. Although, testability is most obviously relevant during testing, but paying attention to testability early in the development process, testing efficiency and effectiveness may potentially be improved. Testability can be perceived as the property and/or characteristic that measures the ease of testing a piece of code or functionality, and a provision added in software so that test plans and scripts can be executed systematically. Testability analysis can add information that is useful both for assessing the overall quality and for locating software bugs [3]. Hence, it provides a trade-off analysis tool for designers to help them in deciding whether they are willing to pay the penalty for testability at the cost of other benefits.

## 3 FACTORIZING TESTABILITY

Testability is now established to be a distinct software quality characteristic. An accurate measure of software quality depends on testability measurement, which in turn depends on the factors that can affect testability. The notion of software testability has been a subject to the number of different interpretations by the experts. However, testability has always been an elusive concept and its correct measurement a difficult exercise [17]. Further, there is not much consensus among practitioners about *'what aspects are actually related to software testability'* [5]. Moreover, there is a conflict among practitioners in considering the factors while estimating testability in general and exclusively at design level. Hence, it is difficult to get a clear view on all the potential factors that can affect testability and the dominant degree of these factors under different testing contexts. Following sections briefly present some of the relevant efforts and contributions made by researchers and practitioners in the direction of finding the testability factors.

Robert Binder, a testability Guru did a novel work in finding out the need and significance of testability for developing software system [4]. He considered six factors of testability including representation characteristics, implementation characteristics, built-in-test capabilities, test suite, test support environment, development process. These factors are described at a high level of abstraction leading to no clear relationship with the metrics that are based on design artifacts and the implementation. Bruce and Haifeng Shi identify the factors that affect testability in object-oriented software and categorized them into structure factors, communication factors and inheritance factors [2]. James Bach describes testability as composed of 'Control, Visibility, Operability, Simplicity, Understandability, Suitability and Stability' [15]. Jungmayr identified software related factors that affect testability at design stage [8]. These testability factors were grouped into nine factors namely, complexity, separation of concerns, coupling, fault locality, controllability, observability, automatability, built-in-test capability and diagnostic capability. Jerry and Ming proposed pentagram model to measure testability, which is based on the factors 'controllability, observability, understandability, traceability, process capability'[6].

The foregoing description reveals that there is a conflict among practitioners in considering the factors while estimating testability in general and exclusively at design level. Therefore, it seems highly desirable to identify the factors affecting testability. Though, getting a universally accepted set of testability factor is impossible, an effort has been made to collect a commonly accepted set of factors that can affect testability. However, without any loss of generality, it appears reasonable to include the factors namely 'complexity, controllability, observability, understandability, traceability and built-in-test' as testability factors.

## 4 OO CHARACTERISTICS

In today's software development environments, object-oriented analysis and design are the popular concepts. They are often heralded as the silver bullet for solving software problems while in reality there is no silver bullet. However, it has proved its value for systems that must be maintained and modified. Object oriented methodologies focus on objects as the primary agents involved in a computation. Each class of data and related operations are collected into a single system entity. It requires much significant effort at the initial stages in the development life cycle to identify objects and classes, attributes and operations and the relationships between them. Object oriented programming is a basis technology, that supports quality goals but just knowing the syntax elements of language and/or the concepts involved in the object oriented technology is far from being sufficient to produce quality software [19].

The necessity to deal with the increasing complexity of software programs is the essential factor that influenced the evolution of programming paradigms. OO programming provides us with a set of proper mechanisms for the management of this complexity. A good object oriented design needs design rules and practices that must be known and used. Their violation will eventually have a strong impact on the quality attributes. Advances in quality and productivity need to be correlated with the use of language constructs. It is then needed to evaluate this use quantitatively to guide OO design. Object oriented principles guide the designers what to support and what to avoid. Several measures have been defined so far to estimate object oriented design. There are several essential themes of object orientation that are known to be the basis of internal quality of object oriented design and support in the context of measurement. These themes prominently include cohesion, coupling, encapsulation and inheritance.

Cohesion refers to the internal consistency within the parts of the design. A class is cohesive when its parts are highly correlated. It should be difficult to split a cohesive class. Cohesion can be used to identify the poorly designed classes. Coupling indicates the relationship or interdependency between modules. For example, object X is coupled to object Y if and only if X sends a message to Y that means the number of collaboration between classes or the number of messages passed between objects. Coupling is a measure of interconnecting among modules in a software structure. Inheritance is the sharing of attributes and operations among classes based on a hierarchical relationship. It is a mechanism whereby one object acquires characteristics from one, or more other objects. Inheritance occurs in all levels of a class hierarchy. Encapsulation is a mechanism to realize data abstraction and information hiding. It hides internal specification of an object and show only external interface. It influences metrics by changing the focus of measurement from a single module to a package of data.

## 5 ESTABLISHING RELATION

In order to establish a contextual impact-relationship between OO software characteristics and testability factors, the influence of OO characteristic on each factor of testability was examined by several researchers. Most of the studies focused their attempt to examine the impact of object oriented characteristics and have successfully established relationships with quality factors. However, we examined and assessed their impact on the particular aspect of study i.e. testability and by associatively and congruence perspective, concluded on identifying testability factors affected by object characteristics. It was observed that each of these characteristics, either have positive or negative impact on the factors that affect testability of object oriented software. After an exhaustive review of available literature on the topic [2], [4], [17], [18], [20], the relation between OO software characteristics and testability factors (as depicted in Figure1) has been established. Based on the relationship shown below, a model has been developed in section 7 (equation 2) for estimating Understandability. Further, the relative significance of individual design properties that influence software testability is weighted proportionally. The concept of multiple linear regressions has been used to get the coefficients that establish a relationship between dependent variables and multiple independent variables.

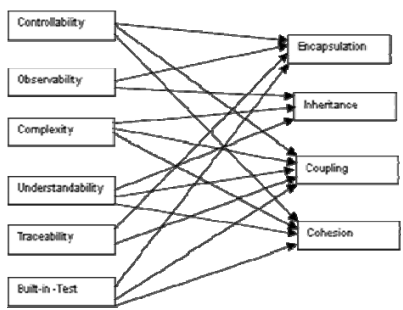

**Figure1: Mapping**

## 6 UNDERSTANDABILITY

Software systems tend to depart more and more from the principle of simplicity and become increasingly complex. The increase in size and complexity of software drastically affects several quality attributes, especially understandability and maintainability. Software developers and maintainers need to read and understand source programs and other documents of software. The significance of understandability is very obvious that can be perceived as 'If we can't learn something, we won't understand it. If we can't understand something, we can't use it - at least not well enough to avoid creating a money pit. We can't maintain a system that we don't understand - at least not easily. And we can't make changes to our system if we can't understand how the system as a whole will work once the changes are made' [16]. Understandability of software documents is thus important as 'the better we know what the thing is supposed to do, the better we can test for it'.

Despite the fact that understandability is vital and highly significant to the software development process, it is poorly managed [1]. The fundamental reality of measurement 'we cannot control what we cannot measure' highlights the importance and significance of good measure of software understandability [13]. There are some aspects of software artifacts that either directly or indirectly affects understandability of software. Classes that are huge and complex are hard to understand by humans, especially if the class is of low cohesion. A good software design with manageable complexity usually provides proper data abstraction; it reduces coupling while increasing cohesion that make them easily understandable. Researchers and Practitioners advocated that understandability aspect of software is highly desirable and significant for developing quality software. Literature survey reveals that there are various aspects of software, including understandability factor that either directly or indirectly influence testability of software [6], [7], [15]. Aforementioned facts reveal that understandability is a key factor to testability.

## 7 MODEL DEVELOPMENT

In order to establish a model for understandability, multiple linear regression technique has been used. Multivariate linear model is given as follows.

$$Y = a_0 + a_1 x_1 + a_2 x_2 + a_3 x_3 \ldots\ldots a_n x_n \quad (1)$$

Where

Y is dependent variable, $x_1, x_2, x_3 \ldots\ldots x_n$ are independent variables related to Y and are expected to explain the variance in Y.
$a_1, a_2, a_3, \ldots\ldots a_n$ are the coefficients of the respective independent variables and $a_0$ is the intercept.

The data used for developing model is taken from [12] that have been collected through the controlled experiment. It includes a set of 28 class diagrams (denoted as $D_0$ to $D_{27}$) and the metrics value of each diagram. Moreover, the mean value of the expert's rating of understandability of these diagrams is also given and termed as 'Known Value' in this paper. The relationship between Testability Factors and Object Oriented Characteristics has been established as depicted in figure1. As per the mapping, Metrics 'NAssoc, NA and MaxDIT' are selected from Table1 as independent variables to be replaced by $x_1$, $x_2$, and $x_3$, respectively to formulate the understandability model. Using MATLAB, values of coefficients are determined and understandability model is formulated as given below.

Understandability = $1.33515 + 0.129 x_1 + 0.0463 x_2 + 0.3405 x_3$ (2)

Table 1: Metrics Description [12]

| Metric | Definition |
|---|---|
| NC | The total number of classes |
| NA | The total number of attributes |
| NM | The total number of methods |
| NAssoc | The total number of associations |
| NAgg | The total number of aggregation relationships within a class diagram (each whole-part pair in an aggregation relationship) |
| NDep | The total number of dependency relationships |
| NGen | The total number of generalization relationships within a class diagram (each parent-child pair in a generalization relationship) |
| NAggH | The total number of aggregation hierarchies in a class diagram |
| NgenH | The total number of generalization hierarchies in a class diagram |
| Max Hagg | It is the maximum between the HAgg values obtained for each class of the class diagram. The HAgg value for a class within an aggregation hierarchy is the longest path from the class to the leaves. |
| Max DIT | It is the maximum between the DIT values obtained for each class of the class diagram. The DIT value for a class within a generalization hierarchy is the longest path from the class to the root of the hierarchy. |

## 8 EXPERIMENTAL TRYOUT

No matter how powerful a theoretical result may be, it has to be empirically validated if it is going to be of any practical use. This is true in all Engineering disciplines, including Software Engineering. Therefore, in addition to the theoretical validation, an experimental tryout is equally important in order to make the claim acceptable. In view of this fact, an experimental validation of the proposed model (equation 2) has been carried out with the help of metrics given in the data set [12]. As per the relationship between OO characteristics and testability factors depicted in figure1, and the understandability Model, the metrics 'NAssoc, NA and MaxDIT' were used to calculate the understandability of design diagrams. Summary of the values obtained the model against the 'Known Values' of understandability are given in Table 2.

Speraman's Rank Correlation coefficient $r_s$ was used to test the significance of correlation between calculated values of understandability using model and it's 'Known Values'. The '$r_s$' was computed using the formula given as under:

$$r_s = 1 - \frac{6 \sum d^2}{n(n^2 - 1)} \qquad -1.0 \leq r_s \leq +1.0$$

Where 'd' is the difference between 'Calculated Values' and 'Known Values' of understandability. And n is the number of UML diagrams (n=28) used in the experiment.

Table 2: Summary of Ratings

| CD | Understandability | |
|---|---|---|
| | Known Value | Value obtained using Model |
| $D_0$ | 1.000 | 1.667 |
| $D_1$ | 2.000 | 1.759 |
| $D_2$ | 2.000 | 1.898 |
| $D_3$ | 2.000 | 2.065 |
| $D_4$ | 2.000 | 2.129 |
| $D_5$ | 2.000 | 1.889 |
| $D_6$ | 2.000 | 2.111 |
| $D_7$ | 3.000 | 2.415 |
| $D_8$ | 2.000 | 1.898 |
| $D_9$ | 3.000 | 2.510 |
| $D_{10}$ | 3.000 | 2.785 |
| $D_{11}$ | 3.000 | 2.915 |
| $D_{12}$ | 3.000 | 3.010 |
| $D_{13}$ | 2.000 | 2.449 |
| $D_{14}$ | 2.000 | 2.928 |
| $D_{15}$ | 4.000 | 3.500 |
| $D_{16}$ | 6.000 | 5.275 |
| $D_{17}$ | 6.000 | 5.569 |
| $D_{18}$ | 6.000 | 5.575 |
| $D_{19}$ | 6.000 | 6.863 |
| $D_{20}$ | 3.000 | 3.118 |
| $D_{21}$ | 5.000 | 4.491 |
| $D_{22}$ | 6.000 | 5.838 |
| $D_{23}$ | 5.000 | 4.665 |
| $D_{24}$ | 5.000 | 5.175 |
| $D_{25}$ | 6.000 | 6.134 |
| $D_{26}$ | 4.000 | 4.806 |
| $D_{27}$ | 4.000 | 4.385 |

Spearman's Rank Correlation Coefficient ($r_s$ = 0.9482) calculated for the proposed model is more than the threshold value for n=28. This shows that the values of understandability computed using model are highly correlated with the 'Known Values'. Thus, the correlation is acceptable with the high degree of confidence, i.e. at the .05. Therefore, without any loss of generality we can con-clude that Understandability Model (equation 2) estimates are reliable and valid in the context. However, the study needs to be standardized with a larger experimental tryout for better acceptability and utility.

## 9 CONCLUSION AND FUTURE WORK

Software understandability is vital and one of the most significant components of the software development. The lack of understandability aspect often leads to false interpretation that may in turn lead to ambiguities, misunderstanding and hence to faulty development results. The paper highlighted the importance of understandability in general and as a factor of software testability in particular. It plays an important role as far as the issue of delivering quality software is concerned. Therefore, Understandability is obviously relevant and significant in the context of software testability. The model has been validated theoretically as well as empirically using experimental try-out. Result shows that the values of understandability computed through model are highly correlated with the 'known values'. However, the model has been validated on a small data set and it is to be done further on live industrial projects for better acceptability and utility.

**Mohd Nazir,** Assistant Professor in Department of Computer Science, Jamia Millia Islamia (Central Unversity), New Delhi, India. He has around six years of teaching experience at PG level. He has obtainned his B. Sc (Hons) & Master of Computer Applications (MCA) degrees both from Aligarh Muslim University, Aligarh, India. He is currently pursuing PhD degree in the area of Software Testing.

**Raees A. Khan, Associate Professor & Head,** Department of Information Technology, BBA University, Lucknow (INDIA) Dr. Raees A. Khan done MCA from Punjab University & Ph.D. (Computer Science) from Jamia Millia Islamia ( Central University), New Delhi. He has written one book and coauthored one book, published numerous articles, several papers in the Nationnal and International Journals and conference proceedings. The area of expertise are Software Quality Assurance, Software testing, Software Security. He has written two projects on Software Security. He is a member of various professional bodies.

**Khurram Mustafa, Professor & Head,** Department of Computer Science, Jamia Millia Islamia (Central Unversity), New Delhi, India. Dr. Mustafa has obtained his M. Tech. & Ph.D. degrees both from IIT Delhi, India. He has written two books, published numerous articles, several papers in the Nationnal and International Journals and conference proceedings. One of his books '*Software Testing: Concepts and Practices*', has been translated into Chinees and published by Science publication. His areas of expertise are Multimedia Applications, Software Quality Assurance, Software Testing, and Software Security. He has around Fourteen Years of teaching experience in India and abroad. He has supervised several projects, including one as a principal investigator on Software Security (funded by DIT, Govt of India). He has contributed a lot in academics & research in the form of delivering lectures, key-note addresses & curriculum developments. He has also organized Seminars, Natinal & International Conferenes. He is a member of various professional bodies.